\documentclass{llncs}

\usepackage{latexsym}
\usepackage{amssymb,amsmath}
\usepackage[group-separator={,}]{siunitx}
\usepackage{nameref}
\usepackage{complexity}
\usepackage{graphicx}
\usepackage{multirow}
\usepackage{caption}

\newtheorem{mytheorem}{Theorem}
\newtheorem{mylemma}{Lemma}  
\newtheorem{myproposition}{Proposition}
\newtheorem{myexample}{Example}
\newtheorem{myobservation}{Observation}

\newcommand{\myqed}{\mbox{$\Box$}}

\newcommand{\mylike}{\textsc{Li\-ke}}
\newcommand{\myblike}{\textsc{Ba\-la\-n\-ced Li\-ke}}
\newcommand{\mymax}{\textsc{Ma\-xi\-mum Li\-ke}}
\newcommand{\myrank}{\textsc{Ra\-nk\-ing}}
\newcommand{\myrand}{\textsc{Ra\-n\-d\-om}}

\newcolumntype{C}{>{\centering\arraybackslash}p{2cm}}
\newcolumntype{L}{>{\centering\arraybackslash}p{2cm}}

\begin{document} 

\title{Most Competitive Mechanisms \\ in Online Fair Division}
\author{Martin Aleksandrov \and Toby Walsh}
\institute{UNSW Sydney and Data61, CSIRO and TU Berlin \email{\{martin.aleksandrov,toby.walsh\}@data61.csiro.au}}
\maketitle

\begin{abstract}
This paper combines two key ingredients for online algorithms - \emph{competitive analysis} (e.g.\ the competitive ratio) and \emph{advice complexity} (e.g.\ the number of advice bits needed to improve online decisions) - in the context of a simple online fair division model where items arrive one by one and are allocated to agents via some mechanism. We consider four such online mechanisms: the popular \myrank\ matching mechanism adapted from online bipartite matching and the \mylike, \myblike\ and \mymax\ allocation mechanisms firstly introduced for online fair division problems. Our first contribution is that we perform a competitive analysis of these mechanisms with respect to the expected size of the matching, the utilitarian welfare, and the egalitarian welfare. We also suppose that an oracle can give a number of advice bits to the mechanisms. Our second contribution is to give several impossibility results; e.g. \emph{no} mechanism can achieve the egalitarian outcome of the optimal offline mechanism supposing they receive partial advice from the oracle. Our third contribution is that we quantify the competitive performance of these four mechanisms w.r.t. the number of oracle requests they can make. We thus present a {\em most competitive} mechanism for each objective.
\end{abstract}

\section{Introduction}\label{sec:intro}

Competitive analysis is a well-known technique to measure the quality of online versus offline decisions \cite{borodin1998,sleator1985}. \emph{Online} decisions are irrevocable (i.e.\ we cannot change past decisions) and instantaneous (i.e.\ we cannot use future knowledge). \emph{Offline} decisions are made supposing the entire problem information is available. Competitive analysis has been applied in various areas during the years, e.g.\ online bipartite matching, online stochastic matching, online sequential allocation, online sequential bin packing, online scheduling \cite{albers2012,gyorgy2010,jaillet2014,karp1990,khuller1994}.

For some online problems, quite successful algorithms are already known under particular assumptions about the arriving input (e.g.\ \cite{brubach2016}). For other problems, this is unfortunately not the case. For example, in the uniform knapsack problem, any deterministic online algorithm without advice has an unbounded competitive ratio. Interestingly, with just one bit of advice, it is possible to implement a 2-competitive algorithm for this problem \cite{marchetti1995}. In general, we can increase the competitive ratio of any online algorithm by giving it enough advice. This motivates the development of novel frameworks such as \emph{advice complexity}. 

An online algorithm has now an access to an \emph{oracle tape} for the problem of interest and can request an \emph{advice string} when making a decision. The oracle is normally assumed to have an unlimited computational power but the number of bits in the advice string must be polynomially bounded in the size of the input offline problem. For a detailed survey on advice complexity, we refer to \cite{boyar2016}. Advice complexity is also related to \emph{semi-online} and \emph{look-ahead} algorithms that suppose some of the input is available \cite{seiden2000}.

This raises a number of questions. 
How many advice bits are sufficient to increase the competitive ratio of an online algorithm to a certain threshold? How many bits are needed to match an {\em optimal} offline algorithm? For example, in the popular paging problem, to achieve offline optimality with an online algorithm we need $\lceil\log_2 k\rceil$ bits of advice to specify which page to delete from the buffer of size $k$. This results in advice complexity of $n\cdot\lceil\log_2 k\rceil$ for instances with $n$ requests, whereas it is shown that $n+k$ bits of advice suffice \cite{bockenhauer2009,dobrev2009}. As another example, in online bipartite matching with a graph of size $n$ (i.e.\ the number of vertices in a partition), a corresponding deterministic online algorithm is optimal (w.r.t.\ the expected matching size) whenever it has an access to $\lceil\log_2 n!\rceil$ but not less advice bits \cite{miyazaki2014}.

Here, for the first time, we introduce techniques from competitive analysis {\em and}  advice complexity into 
online fair division. Online fair division is an important and challenging problem facing society today due to the uncertainty we may have about future resources, e.g.\ deceased organs to patients, donated food to charities, electric vehicles to charging stations, tenants to houses, 
even students to courses, etc. We often cannot wait until the end of the year, week or even 
day before starting to allocate incoming resources. For example, organs cannot be kept too long on ice or products cannot be stored in the warehouse before distributing
to a food bank \cite{walsh2014,walsh2015}. We
extend past work by asking how many advice bits are needed to increase the welfare.

Advice helps us understand how the competitive ratio depends on uncertainty about the future. It can be based on information about past or future items. For example, consider the allocation of food donations to charities by a central decision maker. A number of contractors usually donate food on a regular basis and at specific times so the decision maker knows when some of the items will arrive. Also, each item could have a \emph{type} that is the set of charities that like the item. The oracle might then keep track on the item types that have arrived in the past and thus bias the allocation of the new item type whenever possible. As another example, consider the allocation of deceased organs to patients. The administrator of a hospital might know what organs will arrive that can be exchanged with a neighboring hospital. They might use this offline information to improve significantly the best local online match for the current organ. Further, the oracle could keep track of how long patients are in the waiting list and thus bias the future organ matching decisions based on this information under various constraints, e.g.\ a patient should not wait for an organ more than 30 days, a patient who arrived at time moment 10 to the waiting list should receive organs earlier than a patient who arrived after time moment 10, etc. 

\emph{Our contributions:} Our work is novel for several reasons. For example, we combine advice complexity and competitive analysis in the context of online fair division. As another example, we study multiple objectives and online competitiveness of mechanisms. We first observe a 1-to-1 correspondence between online bipartite matching and online fair division. By using this correspondence, we can transfer and significantly extend objectives and algorithms from online bipartite matching to online fair division and vice versa. This is useful for a number of reasons. For example, agents in fair division have preferences and can be strategic which is an aspect not typically considered in bipartite matching. As a second example, allocations may be more difficult to find than matchings if we want them to satisfy multiple fairness and efficiency criteria. We thus view algorithms for online bipartite matching as \emph{mechanisms} for online fair division. Following this, we study the competitive performance of the popular matching \myrank\ mechanism and the attractive \mylike, \myblike\ and \mymax\ allocation mechanisms w.r.t. three different objectives: the \emph{expected matching size}, the \emph{utilitarian welfare} and the \emph{egalitarian welfare}. We consider three settings, namely online fair division setting \emph{without advice}, \emph{with full advice} and \emph{with partial advice}. In each of these settings, we analyse these four mechanisms and present a most competitive mechanism for each objective supposing adversarial input. We further plot their competitive ratios. We finally proved that there is no mechanism that maximizes the expected matching size or the egalitarian welfare and uses less than full advice.

The next Section~\ref{sec:pre} provides the notions, the mechanisms and the objectives that we use throughout the paper. In Sections~\ref{sec:on},~\ref{sec:advone} and~\ref{sec:advtwo}, we report our results for the online setting without advice, the online setting with full advice and the online setting with partial advice, respectively. Finally, we discuss related work, future work and conclude in Section~\ref{sec:rel}. 

\section{Preliminaries}\label{sec:pre}

{\bf Online bipartite matching instance:} An \emph{instance} $\mathcal{G}$ has (1) a set of $n$ ``\emph{boy}'' vertices, (2) a set of $m$ ``\emph{girl}'' vertices, (3) a \emph{weight} matrix where the $(i,j)$-th cell contains the \emph{weight} of the edge between vertices the $i$-th ``boy'' vertex and the $j$-th ``girl'' vertex, and (4) a \emph{sequence} of the ``girl'' vertices. We consider \emph{binary} (i.e.\ unweighted graph) and \emph{non-negative} (i.e.\ weighted graph) weights.

{\bf Online fair division instance:} An \emph{instance} $\mathcal{I}$ has (1) a set $A$ of \emph{agents} $a_1,\ldots,a_n$, (2) a set $O$ of indivisible \emph{items} $o_1,\ldots,o_m$, (3) \emph{utility} matrix $U=(u_{ij})_{n\times m}$ where $u_{ij}$ is the private \emph{utility} of agent $a_i$ for item $o_j$, and (4) \emph{ordering} $o$ of the items. We consider \emph{binary} and \emph{general} non-negative rational utilities. 

{\bf Online setting:} Let $\mathcal{G}_{\mathcal{I}}=(A,O,U,o)$ be the online bipartite graph associated with $\mathcal{I}$. We suppose that ordering $o$ reveals item $o_j$ in round $j$ when each agent $a_i$ bids a rational non-negative value $v_{ij}$ for item $o_j$ and a \emph{mechanism} allocates item $o_j$ to an agent. Further, we assume that the ordering $o$ is adversarial which captures the worst-case arrival sequences. 

{\bf Fair division axioms:} A mechanism is \emph{strategy-proof} if, with complete information, no agent can misreport their utilities and thus increase their expected outcome. Agent $a_1$ \emph{envies (ex ante) ex post} agent $a_2$ if $a_1$ assigns greater (expected) utility to the (expected) allocation of $a_2$ than to their own (expected) allocation. A mechanism is \emph{bounded envy-free (ex ante) ex post with $r$} if no agent envies (ex ante) ex post another one with more than $r$ given the (expected) allocation returned by the mechanism. A mechanism is \emph{(ex ante) ex post Pareto efficient} if its returned (expected) allocation is Pareto optimal.

{\bf Mechanisms:} We use an oracle tape to specify some of the behavior of the optimal offline mechanism. An \emph{online} mechanism $M$ does not consult the oracle tape and makes the current decision supposing the past decisions are irrevocable and no information about future items is available. By comparison, its modification {\sc Adviced} $M$ can at each round decide whether to consult the oracle or not. If ``yes'', the oracle encodes the identifier of the agent that should receive the item on the advice tape when the mechanism reads the tape and allocates the item to the adviced agent. If ``no'', {\sc Adviced} $M$ runs $M$ to allocate the current item. There are two extreme cases. If {\sc Adviced} $M$ does not read the oracle tape at any round, then its performance coincides with the one of $M$. If {\sc Adviced} $M$ reads the oracle tape at each round, then its performance coincides with the one of an optimal offline mechanism. 

We consider four online mechanisms. The \mymax\ mechanism allocates each item $o_j$ uniformly at random to an agent with the greatest bid for $o_j$. The \myrank\ mechanism from \cite{karp1990} picks a strict priority ordering over the agents uniformly at random and allocates each item $o_j$ to an agent that has positive bid for it, has not been allocated items previously and has the greatest priority. We further use the \mylike\ and \myblike\ mechanisms from \cite{aleksandrov2015ijcai}. The \mylike\ mechanism allocates each item $o_j$ uniformly at random to an agent that bids positively for the item. The \myblike\ mechanism allocates each item $o_j$ uniformly at random to an agent among those agents who bid positively for the item and have been allocated fewest items previously. We modify these four mechanisms to read advice bits from the oracle tape: {\sc Adviced} \mymax, {\sc Adviced} \myrank, {\sc Adviced} \mylike\ and {\sc Adviced} \myblike.

These mechanisms satisfy many nice axioms. For example, \mymax\ is Pareto efficient. In fact, it is one of a few Pareto efficient mechanisms but unfortunately it is not strategy-proof or envy-free. \mylike\ is strategy-proof and envy-free ex ante. In fact, each envy-free ex ante mechanism assigns probabilities for items to agents as \mylike\ does. However, \mylike\ is not envy-free ex post. In contrast, \myblike\ mechanism bounds the envy ex post. Interestingly, with 0/1 utilities, it is also Pareto efficient and envy-free ex ante. We further analysed the matching \myrank\ mechanism from a fair division point of view. For example, it is strategy-proof, envy-free ex ante and bounds the envy ex post but only with simple 0/1 utilities. However, it is not Pareto efficient in this setting as it may discard items. These axiomatic properties are well-understood. We, therefore, turn attention to the competitive properties of these mechanisms.  

{\bf Objectives:} Given instance $\mathcal{I}$, each mechanism induces a probability distribution over a set $\Pi_{\mathcal{I}}$ of allocations. The \emph{expected matching size} $\overline{k}(\mathcal{I})$ is equal to $\sum_{\pi\in \Pi_{\mathcal{I}}} p(\pi)\cdot k(\pi)$ where $p(\pi)$ is the probability of allocation $\pi$ and $k(\pi)$ is the number of agents that are allocated items in $\pi$. The \emph{expected utility} $\overline{u}_{i}(\mathcal{I})$ of agent $a_i$ is $\sum_{j=1}^m p_i(j,\mathcal{I})\cdot u_{ij}$ where $p_i(j,\mathcal{I})$ is the expected probability of agent $a_i$ for item $o_j$. The \emph{utilitarian welfare} $\overline{u}(\mathcal{I})$ is equal to $\sum_{i=1}^n \overline{u}_{i}(\mathcal{I})$. The \emph{egalitarian welfare} $\overline{e}(\mathcal{I})$ is equal to $\min_{i=1}^n \overline{u}_{i}(\mathcal{I})$.

\begin{myexample}\label{exp:one} $(${\bf Upper-triangular instance}$)$
Consider $\mathcal{I}$ with $n$ agents, $n$ items and let each agent $a_i$ has utilities equal to 1 for items $o_1$ to $o_{n-i+1}$. For a deterministic mechanism that allocates all items to agents that like them, we have that $\overline{k}(\mathcal{I})\in\lbrace 1,2,\ldots,n\rbrace$, $\overline{u}(\mathcal{I})=n$ and $\overline{e}(\mathcal{I})\in\lbrace 0,1\rbrace$. \hfill\myqed
\end{myexample}

{\bf Performance measures:} We use the objectives in order to define three statistics to measure the performance of online mechanisms over all instances.

\begin{equation}
(ES) \min_{\mathcal{I}}\overline{k}(\mathcal{I})
\end{equation}
\begin{equation}
(UW) \min_{\mathcal{I}}\overline{u}(\mathcal{I})
\end{equation}
\begin{equation}
(EW) \min_{\mathcal{I}}\overline{e}(\mathcal{I})
\end{equation}

{\bf Online ratios with advice:} We say that an online mechanism $M$ has an \emph{offline (online) competitive ratio} $c(m)$ with $m$ advice bits w.r.t. welfare $W$ if, for an instance $\mathcal{I}$ and an ordering $o$ of $m$ items, we have that $W(OPT_{off(on)})\leq c(m)\cdot W(M(\mathcal{I}))+b(m)$ holds where $b(m)$ is an additive constant and $OPT_{off(on)}$ is the optimal offline (online) mechanism. Note that the $OPT_{off}$ mechanism does not depend on the ordering of the items whilst $OPT_{on}$ does. A mechanism $M$ is \emph{most $c(m)$-competitive} w.r.t. welfare $W$ if $M$ has a competitive ratio $c(m)$ w.r.t. $W$ and each other mechanism has a competitive ratio that is at least $c(m)$. We say that $M_1$ is \emph{strictly better} than $M_2$ on a set of instances if the welfare value of $M_1$ is not lower than the one of $M_2$ on all instances from the set, and greater than the one of $M_2$ on some instances from the set. We say that $M_1$ and $M_2$ are \emph{incomparable} if $M_1$ is strictly better than $M_2$ on some instances and $M_2$ is strictly better than $M_1$ on some other instances.

We suppose throughout the paper that agents \emph{sincerely} report their utilities for items. Also, we assume that each agent has positive utility for at least one item and each item is liked by at least one agent. We show our results for the case when $m=n$ and there is a \emph{perfect allocation} in $\mathcal{I}$ (or \emph{perfect matching} in $\mathcal{G}_{\mathcal{I}}$), i.e.\ an allocation in which each agent receives exactly one item that they like. However, we also draw conclusions for the case when $m>n$ and there is an allocation in which each agent receives at least one item that they like. Finally, we extended all our results to the case when the maximum number of agents that receive items that they like in each possible allocation is $k<n$ (or \emph{maximum matching} in $\mathcal{G}_{\mathcal{I}}$ of cardinality $k<n$). However, we omit these results for reasons of space. 

\section{Online Fair Division without Advice}\label{sec:on}

We study the competitiveness of our four online mechanisms w.r.t. to the optimal offline mechanism for the expected matching size (ES), the utilitarian welfare (UW) and the egalitarian welfare (EW). The optimal \emph{offline} mechanism returns an allocation in which each agent receives exactly one item for (ES), an allocation in which each item is received by an agent that values it most for (UW) and a perfect allocation that maximizes the egalitarian welfare for (EW).

\subsection{Expected Matching Size}\label{subsec:exp}

A mechanism that maximizes the objective $\overline{k}(\mathcal{I})$ also maximizes both $\overline{u}(\mathcal{I})$ and $\overline{e}(\mathcal{I})$ simultaneously when agents have simple binary utilities. By Theorem 2 from \cite{miyazaki2014}, no deterministic online mechanism can maximize (ES). We, therefore, turn our attention to randomized mechanisms for (ES). By \cite{karp1990}, the \myrank\ mechanism is most competitive for (ES) with expected matching size of $n\cdot (1-\frac{1}{e})+o(n)$ when the arriving sequence is adversarial. Its competitive ratio is $1+\frac{1}{e-1}$. For this reason, we next report the competitive ratios of \myblike, \mylike\ and \mymax\ with respect to the optimal offline mechanism and \myrank. The optimal offline mechanism returns a matching of expected size $n$.

\begin{mytheorem}\label{thm:one}
The \mylike\ and \myblike\ mechanisms are $2$-competitive and $2\cdot(1-\frac{1}{e})$-online competitive whereas the \mymax\ mechanism is $n$-competitive and $n\cdot(1-\frac{1}{e})$-online competitive for (ES).
\end{mytheorem}

\begin{myproof}
For \myblike, consider the \myrand\ mechanism that allocates each next item uniformly at random to an agent among those with 0 items. If no such agent exists for the current item, then \myrand\ discards the item. The \myblike\ mechanism can be seen as a \emph{completion} of \myrand, i.e.\ \myblike\ allocates even the items that \myrand\ discards. It is easy to prove that the expected matching sizes of \myblike\ and \myrand\ are the same for each instance. Therefore, the \myblike\ and \myrand\ mechanisms achieve the same expected matching size for each instance. By \cite{karp1990}, the minimum such size is equal to $\frac{n}{2}+o(\log_2 n)$.

For \mylike, consider $n$ agents, $n$ items. Suppose that each agent likes the first $n/2$ items. The remaining $n/2$ items are chosen by the adversary. We have that $k\in[1,n/2]$ different agents are allocated the first $n/2$ items. The adversary then chooses the next $n/2$ items in such a way so that $n/2$ different agents like them and $k$ of them are the ones matched the first $n/2$ items. The expected matching size is $\frac{n}{2}+o(\log_2 n)$. There could be instances, however, when this size is lower.

For \mymax, consider an instance with $n$ agents and $n$ items. Let us suppose that all agents have positive utilities for all items but only agent $a_1$ has the greatest utility for each item. The mechanism gives all items to agent $a_1$ and thus achieves a matching size of 1. Note that this is the worst possible outcome. For each instance, the expected matching size of this mechanism is at least 1 because it allocates all items to at least one agent.\hfill \myqed
\end{myproof}

\begin{myobservation}\label{obs:one}
The \myrank\ mechanism is strictly better than the \myblike\ mechanism which is strictly better than the \mylike\ mechanism for (ES).
\end{myobservation}

For \myrank\ and \myblike, the result in Observation~\ref{obs:one} follows immediately from Theorem~\ref{thm:one}. By Lemma~\ref{lem:one}, \myblike\ is at least as competitive than \mylike\ for each instance. For some instances, \myblike\ is more competitive than \mylike. Hence, \myblike\ is strictly better than \mylike. 

\begin{mylemma}\label{lem:one}
Let $\pi_j$ be an allocation of items $o_1$ to $o_j$, and $\rho_j$ and $\sigma_j$ extend $\pi_j$ to all items by using \myblike\ and \mylike, respectively. Further, let $b(\rho_j)$ and $l(\sigma_j)$ be their probabilities. For each instance, $j\in[1,n]$ and $\pi_j$, we have that $\sum_{\rho_j} b(\rho_j)\cdot k(\rho_j)\geq \sum_{\sigma_j} l(\sigma_j)\cdot k(\sigma_j)$ holds.
\end{mylemma}

\myrank\ outperforms \mymax\ in general over all instances. In contrast, there are instances on which \mymax\ outperforms \myrank. We illustrate this in Example~\ref{exp:two}.

\begin{myexample}\label{exp:two} $(${\bf Expected matching incomparabilities}$)$
Consider the fair division of 2 items between 2 agents. Let $u_{11}=2,u_{12}=0,u_{21}=1,u_{22}=2$. The expected matching sizes of \mymax\ and \myrank\ are $2$ and $3/2$. \hfill\myqed
\end{myexample}

If $m>n$, our results hold as well. We conclude that \myrank\ is more competitive than \myblike, \mylike\ and \mymax\ for (ES) in the worst case. 

\subsection{Utilitarian Welfare}\label{subsec:util}

In general, the utilitarian welfare can be maximized even online with no information about future items. One most competitive online mechanism that achieves the optimal offline welfare is \mymax. Hence, the offline and online competitive ratios of online mechanisms conflate to just one competitive ratio.

\begin{myproposition}\label{prop:one}
With general utilities, the \mymax\ mechanism maximizes (UW).
\end{myproposition}

\begin{myproof}
\mymax\ allocates each next item in the ordering to an agent with the greatest utility for the item. The returned online welfare value coincides with the maximum possible offline value of this welfare, i.e.\ the maximum utility sum over the items. \hfill\myqed
\end{myproof}

The result in Proposition~\ref{prop:one} is straightforward in our setting but there are fair division settings in which optimizing the utilitarian welfare is intractable even \emph{offline} when the entire problem input information is available \cite{nguyen2014}. We, therefore, find our result fundamental. On the other hand, with binary utilities, note that each mechanism that gives all items to agents that like them maximizes the utilitarian welfare. Indeed, \myblike\ and \mylike\ do maximize it whereas \myrank\ does not because it might discard items. 

\begin{myobservation}\label{obs:two}
With 0/1 utilities, the \myblike\ and \mylike\ mechanisms are strictly better than the \myrank\ mechanism for (UW).
\end{myobservation}

With general utilities, \mylike\ is $n$-competitive; see the example in the proof of Theorem 9 from \cite{aleksandrov2015ijcai}. By comparison, \myrank\ and \myblike\ are not competitive from a utilitarian perspective even with just two agents and two items. We illustrate these results in Example~\ref{exp:three}.

\begin{myexample}\label{exp:three} $(${\bf Utilitarian non-competitiveness}$)$
Consider the fair division of 2 items to 2 agents. Let $u_{11}=0,u_{12}=1,u_{21}=1,u_{22}=u$. The optimal offline utilitarian welfare is $u+1$ whereas the one of \myblike\ and \myrank\ is $2$. Their ratios go to $\infty$ as $u$ goes to $\infty$. \hfill\myqed
\end{myexample}

Our Example~\ref{exp:three} is in-line with an impossibility example and an impossibility remark presented by \cite{khuller1994} for online weighted bipartite matching. 
These show that there is no deterministic or randomized online algorithm that maximizes (or minimizes) the \emph{perfect utilitarian welfare} (the sum of the utilities in a perfect allocation) where the competitive ratio of the algorithm depends only on the number of agents $n$. In contrast, our utilitarian welfare objective (UW) is different because its maximum value could be obtained by allocating all items to a single agent. As a result, \mymax\ is a mechanism whose competitive ratio does not depend even on $n$ and \mylike\ is a mechanism whose competitive ratio depends solely on $n$. 

If $m>n$, the \mymax\ mechanism remains optimal for (UW). We used the argument in the proof of Theorem 9 from \cite{aleksandrov2015ijcai} to construct an example and show that \mylike\ remains $n$-competitive. Both \myrank\ and \myblike\ remain not competitive; see the example in the proof of Theorem 10 from \cite{aleksandrov2015ijcai}. We conclude that \mymax\ is more competitive than \myrank, \myblike\ and \mylike\ for (UW) in any case. 

\subsection{Egalitarian Welfare}\label{subsec:egal}

In this section, we optimize the egalitarian welfare. It is easy to see that there is no deterministic online mechanism that maximizes the egalitarian welfare. We focus therefore on randomized mechanisms.

With binary utilities, both \mylike\ and \myblike\ are $n$-competitive from an egalitarian perspective; see Example~\ref{exp:one}. Moreover, \mymax\ is equivalent to \mylike\ and hence it is also $n$-competitive. With general utilities, \mymax\ is unfortunately not competitive at all even with just two agents and two items. See Example~\ref{exp:four} for this simple result.

\begin{myexample}\label{exp:four} $(${\bf Egalitarian non-competitiveness}$)$
Consider the fair division of 2 items to 2 agents. Let $u_{11}=2,u_{12}=2,u_{21}=1,u_{22}=1$. An optimal offline egalitarian mechanism gives say item $o_1$ to agent $a_1$ with probability $1$ and item $o_2$ to agent $a_2$ with probability $1$. Its egalitarian welfare is equal to $1$. \mymax\ gives items $o_1$ and $o_2$ to agent $a_1$ with probability 1. Its welfare is equal to 0. Hence, its ratio is $\infty$. \hfill\myqed
\end{myexample}

Interestingly, with general utilities, \mylike, \myblike\ and \myrank\ are all most $n$-competitive from an egalitarian perspective.

\begin{mytheorem}\label{thm:two}
With general utilities, the \myblike, \mylike\ and \myrank\ mechanisms are most $n$-competitive for (EW).
\end{mytheorem}

\begin{myproof}
The mechanisms have competitive ratios of $n$. Consider instance $\mathcal{I}$, agent $a_i$ and the first item $o_j$ in the ordering such that agent $a_i$ has positive utility for it. We show that $\overline{e}(\mathcal{I})$ is at least $\frac{1}{n}$. With \mylike, we have that the probability $p_i(j,\mathcal{I})$ of agent $a_i$ for item $o_j$ is equal to $1/n_j$ where $n_j$ is the number of agents that like item $o_j$. Since $n_j\leq n$, we have $p_i(j,\mathcal{I})\geq 1/n$. With \myblike\ and \myrank, the worst case for agent $a_i$ is when they have been allocated 0 items prior round $j$ and all agents together have positive utilities for item $o_j$. Therefore, we have $p_i(j,\mathcal{I})\geq 1/n_j\geq 1/n$. Hence, agent $a_i$ receives expected utility of at least $\frac{1}{n}$. This lower bound is achieved in Example~\ref{exp:one}.

Next, we confirm that every other mechanism has competitive ratio at least $n$.
Consider the upper-triangular instance from Example~\ref{exp:one} and a mechanism $M$. If $M$ shares the probability for the first item uniformly at random, then its competitive ratio is equal to $\frac{1}{n}$. If $M$ shares the probability for the first item not uniformly at random, then its competitive ratio is lower than $\frac{1}{n}$. Suppose that $M$ gives the first item to agent $a_n$ with probability $p>\frac{1}{n}$. The probability of some other agent must be smaller than $\frac{1}{n}$ as these probability values sum up to at most 1. WLOG, suppose that the probability $q$ of agent $a_1$ for this first item is one such value smaller than $\frac{1}{n}$. The egalitarian welfare on the upper-triangular instance is then $p$. However, consider next the \emph{lower-triangular} instance, i.e.\ agent $a_i$ likes items $o_1$ to $o_i$. The mechanism gives expected utility of $q<\frac{1}{n}$ to agent $a_1$. This value is also the welfare on the lower-triangular instance. $M$ has competitive ratio of $1/q$ because the optimal offline welfare is 1.
\hfill\myqed
\end{myproof}

\begin{myobservation}\label{obs:three}
With 0/1 utilities, the \myrank\ mechanism is strictly better than the \myblike\ mechanism which is strictly better than the \mylike\ mechanism for (EW).
\end{myobservation}

Observation~\ref{obs:three} can be shown similarly as Observation~\ref{obs:one}. Surprisingly, there are instances on which \mymax\ outperforms all the other three mechanisms even though it is not competitive in general. See Example~\ref{exp:five} for this result.

\begin{myexample}\label{exp:five} $(${\bf Egalitarian incomparabilities}$)$
Let $\mathcal{I}$ has 2 items, 2 agents and $u_{11}=2,u_{12}=1,u_{21}=1,u_{22}=2$. The value of $\overline{e}(\mathcal{I})$ of \mymax\ is $2$ whereas the value of $\overline{e}(\mathcal{I})$ of \myblike, \mylike\ or \myrank\ is equal to $3/2$.\hfill\myqed
\end{myexample}

If $m>n$, \myrank\ and \myblike\ become not competitive; see the example in the proof of Theorem 10 from \cite{aleksandrov2015ijcai}. \mylike\ however remains most $n$-competitive; see the example in the proof of Theorem 9 from \cite{aleksandrov2015ijcai}. We conclude that \mylike\ is more competitive than \myrank, \myblike, \mymax\ for (EW) in the worst case. 

\section{Online Fair Division with Full Advice}\label{sec:advone}

We next study most competitive adviced mechanisms for the expected matching size (ES), the utilitarian welfare (UW) and the egalitarian welfare (EW). By Proposition~\ref{prop:one}, there is a deterministic online mechanism that maximizes (UW) even without any advice. We, therefore, focus on (ES) and (EW). 

We assume that the oracle specifies on the tape a different agent for each of the $n$ items. Such an encoding requires $\lceil\log_2 n!\rceil$ advice bits. By Theorem 1 from \cite{miyazaki2014}, there is a deterministic online mechanism that uses $\lceil\log_2 n!\rceil$ advice bits and maximizes (ES). By Theorem 2 from \cite{miyazaki2014}, no deterministic online mechanism can use less than $\lceil\log_2 n!\rceil$ advice bits and maximize (ES). These two results are inherited for (EW) as well. Interestingly, we next prove that no randomized online mechanism can use less than $\lceil\log_2 n!\rceil$ advice bits and maximize either objective (ES) or (EW).

\begin{mytheorem}\label{thm:three}
There is {\bf no} randomized online algorithm that uses less than $\lceil\log_2 n!\rceil$ advice bits and maximizes (ES). Even with 0/1 utilities, there is {\bf no} randomized online algorithm that uses less than $\lceil\log_2 n!\rceil$ advice bits and maximizes (EW).
\end{mytheorem}

\begin{myproof}
For (ES), suppose that there is such a mechanism. The maximum value of (ES) is $n$. Let $\pi$ be an allocation returned by the mechanism and $p(\pi)$ its probability. Recall that $k(\pi)\leq n$ denotes the number of different agents that receive items in $\pi$. If $\sum_{\pi} p(\pi)<1$ holds, then we conclude that $\sum_{\pi} p(\pi)\cdot k(\pi)<n$ holds. Therefore, the mechanism does not maximize (ES) which is a contradiction. Consequently, $\sum_{\pi} p(\pi)=1$ holds. But, now we have that $\sum_{\pi} p(\pi)\cdot k(\pi)<n$ iff $k(\pi)<n$ for some $\pi$ returned by the mechanism. Therefore, as the mechanism maximizes (ES), we conclude that $k(\pi)=n$ for each $\pi$. To sum up, the mechanism returns only perfect allocations and their probabilities sum up to 1. We can define now a deterministic online mechanism given one $\pi$ returned by the randomized online mechanism. This deterministic online mechanism also uses less than $\lceil\log_2 n!\rceil$ advice bits and maximizes (ES). This is in contradiction with Theorem 2 from \cite{miyazaki2014}. This result holds even with more items than agents.

For (EW) and binary utilities, suppose that there is such a mechanism. Hence, each agent receives an expected utility of 1 and the probability of 1 for each item is shared completely between agents that like the item. Given instance $\mathcal{I}$, consider the \emph{random assignment} matrix $P(\mathcal{I})=(p_i(j,\mathcal{I}))_{n\times n}$ of this mechanism. The matrix $P(\mathcal{I})$ is \emph{bistochastic} because $\sum_{i=1}^n p_i(j,\mathcal{I})=1$ for each $j$ and $\sum_{j=1}^n p_i(j,\mathcal{I})=1$ for each $i$ hold. By the famous result of Birkhoff, every bistochastic matrix is a convex combination of permutation matrices \cite{brualdi2006}. Each permutation matrix corresponds to a perfect allocation in $\mathcal{I}$. There could be multiple combinations for the same bistochastic matrix. For each such combination, we can define a randomized online algorithm that uses less than $\lceil\log_2 n!\rceil$ advice bits and maximizes (ES). This is in contradiction with the previous result. This result holds even with more items than agents.\hfill\myqed
\end{myproof}

\section{Online Fair Division with Partial Advice}\label{sec:advtwo}

In this section, we report the reciprocal ratios of the mechanisms. We assume that the oracle specifies agents for $k<m$ items. We start with the case when $m=n$. For (ES), the oracle specifies a different agent for each of the first $k$ items. An efficient encoding requires $\lceil\log_2 k!\rceil$ advice bits. If $k=n-1$, {\sc Adviced} \myrank\ and {\sc Adviced} \myblike\ are optimal because they keep track on the past allocation whereas {\sc Adviced} \mymax\ and {\sc Adviced} \mylike\ have ratios $1-\frac{1}{n}$ and $1-\frac{1}{n}+\frac{1}{n^2}$. If $k<n-1$, we next report their ratios. 

\begin{mytheorem}\label{thm:four}
With $\lceil\log_2 k!\rceil$ advice bits, {\sc Adviced} \myrank\ is most $\frac{(e-1)n+k}{en}$-competitive for (ES).
\end{mytheorem}

\begin{myproof}
The mechanism has two components: (1) one that allocates items deterministically and (2) another one that allocates items according to \myrank. Let the entire input graph be $\mathcal{G}_{\mathcal{I}}$ with $n$ vertices in each partition. Let us remove the $k$ deterministically decided vertices from both partitions together with their edges from $\mathcal{G}_{\mathcal{I}}$. Now, consider the remaining bipartite sub-graph with $(n-k)$ vertices in each partition. This graph has perfect matching of size $(n-k)$ and \myrank\ matches vertices in this graph. Therefore, the expected matching size of \myrank\ on this smaller graph is $(n-k)\cdot(1-\frac{1}{e})+o(n-k)$. We conclude that this size for {\sc Adviced} \myrank\ is $k+(n-k)\cdot(1-\frac{1}{e})+o(n-k)$.

By Theorem 1 from \cite{miyazaki2014}, the deterministic component of {\sc Adviced} \myrank\ maximizes (ES) on the bipartite sub-graph of $\mathcal{G}_{\mathcal{I}}$ that contains the adviced $2\cdot k$ vertices. By \cite{karp1990}, we conclude that the randomized component of {\sc Adviced} \myrank\ maximizes (ES) on the bipartite sub-graph of $\mathcal{G}_{\mathcal{I}}$ that contains the remaining unadviced $2\cdot(n-k)$ vertices.\hfill\myqed
\end{myproof}

\begin{figure}[h]
\resizebox{\textwidth}{!}{
\includegraphics[height=3.75cm,width=0.475\textwidth,clip=true,trim=10 0 35 15]{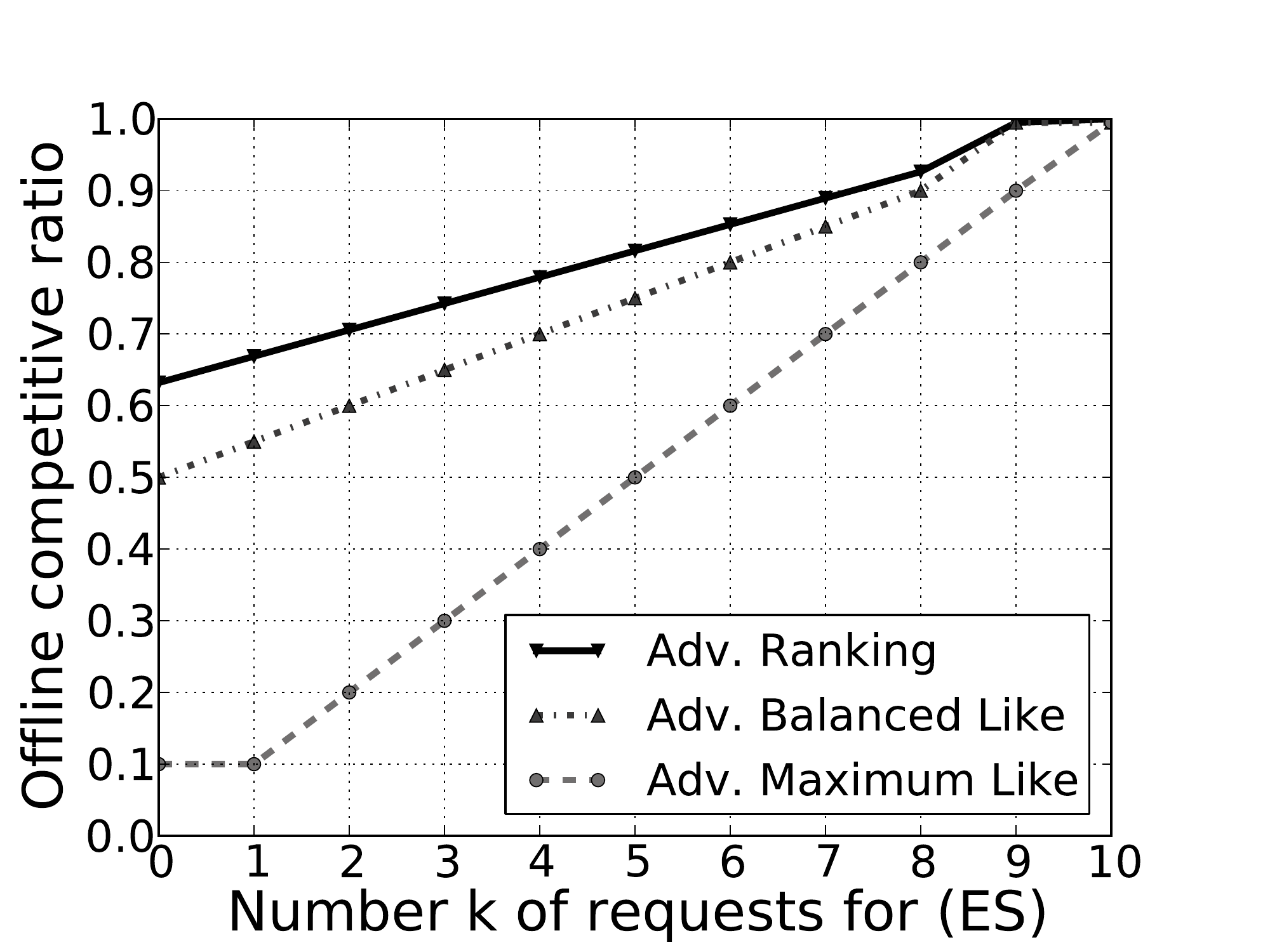}%
\includegraphics[height=3.75cm,width=0.475\textwidth,clip=true,trim=10 0 35 15]{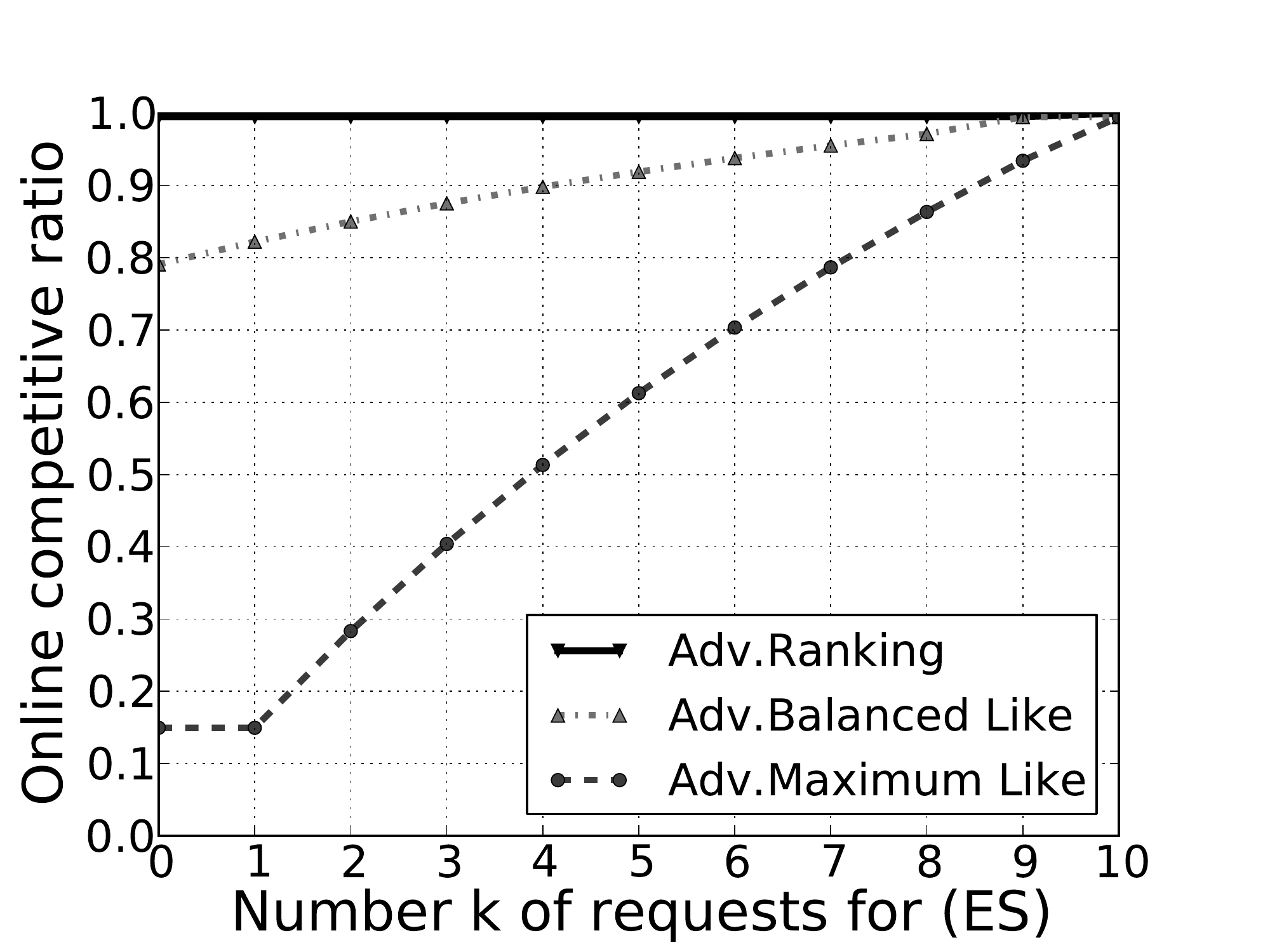}%
}
\caption{(left) w.r.t optimal offline mechanism, (right) w.r.t. {\sc Adviced} \myrank}
\label{fig:one}
\end{figure}

By Theorem 2 from \cite{miyazaki2014} and our Theorem~\ref{thm:three}, there is no mechanism that uses less than $\lceil\log_2 k!\rceil$ advice bits and has a greater competitive ratio than {\sc Adviced} \myrank\ with $\lceil\log_2 k!\rceil$ advice bits. We also obtained that the offline ratios of {\sc Adviced} \mymax, {\sc Adviced} \myblike\ and {\sc Adviced} \mylike\ for (ES) and $k\in[1,n-1)$ are $\frac{k}{n}$, $\frac{k+n}{2n}$ and at most $\frac{k+n}{2n}$. Their online ratios are $\frac{ek}{(e-1)n+k}$, $\frac{e(k+n)}{2(e-1)n+2k}$ and at most $\frac{e(k+n)}{2(e-1)n+2k}$. In Figure~\ref{fig:one}, we plot these ratios for $n=10$ agents and $k\in[0,n]$ oracle calls.        

\vspace{-0.75cm}

For (UW), (EW) and 0/1 utilities, the oracle specifies a different agent for each of the first $k$ items. For (UW) and general utilities, the oracle specifies an agent for each of $k$ most valued items. The worst case for {\sc Adviced} \myrank\ and {\sc Adviced} \myblike\ is when the adviced allocation biases the allocation of future items towards agents who receive negligibly small utilities for these items. Instead, {\sc Adviced} \mylike\ allocates each such unadviced item to an agent with probability at least $\frac{1}{n}$. {\sc Adviced} \mymax\ optimizes (UW) by Proposition~\ref{prop:one}. For (EW) and general utilities, the oracle computes an allocation of $k$ items to agents that maximizes the egalitarian welfare and then specifies the $k$ agents for the $k$ items in this computed allocation. {\sc Adviced} \myrank\ and {\sc Adviced} \myblike\ focus on agents with zero and fewest items whereas {\sc Adviced} \mylike\ and {\sc Adviced} \mymax\ perform as \mylike\ and \mymax.

We next consider the case when $m>n$. For (ES), we conclude the same results as above. For (UW), (EW) and 0/1 utilities, we assume that the oracle specifies $k$ agents for the first $k$ items in the ordering for which the $k$ agents are different. For (UW), (EW) and general utilities, the oracle specifications are as in the case when $m=n$. We summarize all ratios in Table~\ref{tab:one}. 

\vspace{-0.25cm}

\begin{table}[h]
\captionsetup{justification=centering}
\caption{Ratios for $k\in[0,m)$ adviced items and $l\in[1,n)$ adviced agents:\hspace{1cm} (b) - binary utilities, (g)-general utilities.}
\resizebox{\textwidth}{!}{
\begin{tabular}{|C|L|L|L|L|L|}
\hline 
\multirow{2}{1.5cm}{\bf Mechanism} & {\bf (UW)-b} & {\bf (UW)-g} & {\bf (EW)-b} & {\bf (EW)-g} & {\bf (EW)-g} \\

  & $m\geq n$ & $m\geq n$ & $m\geq n$ & $m=n$ &  $m>n$ \\ \hline
 
\multicolumn{1}{|c|}{\sc Adv.Max.Like} & $1$ & $1$ & $\frac{1}{n}$ & $0$ & $0$ \\
\multicolumn{1}{|c|}{\sc Adv.Bal.Like} & $1$ & $\frac{k}{m}$ & $\frac{1}{n-l}$ & $\frac{1}{n-l}$  & $0$ \\
\multicolumn{1}{|c|}{\sc Adv.Like} & $1$ & $\frac{k}{m}+\frac{1}{n}-\frac{k}{nm}$ & $\frac{1}{n}$ & $\frac{1}{n}$ & $\frac{1}{n}$ \\
\multicolumn{1}{|c|}{\sc Adv.Ranking} & $\leq \frac{n}{m}$ & $\frac{k}{m}$ & $\frac{1}{n-l}$ & $\frac{1}{n-l}$  & $0$ \\ \hline
\end{tabular}
}
\label{tab:one}
\end{table}

\vspace{-0.75cm}

\section{Related Work and Conclusions}\label{sec:rel}

We combined competitive analysis, advice complexity and online fair division. Our results are simple but fundamental to understand better the interface between matching and fair division problems. In conclusion, the chair might use {\sc Adviced} \myrank\ for (ES), {\sc Adviced} \mymax\ for (UW) and {\sc Adviced} \mylike\ or {\sc Adviced} \myblike\ for (EW). We quantify the offline and online performance of these mechanisms with respect to the number of advice bits they can read from an oracle tape. We also presented two impossibility results and closed an open question from \cite{miyazaki2014}. 

In future, we will analyse other $b$-matching mechanisms from a fair division viewpoint \cite{jaillet2014,kalyanasundaram2000}. Also, we can explore more objectives (e.g.\ the Nash welfare) or competitive measures (e.g.\ price of anarchy) \cite{aleksandrov2015ijcai,vincent2016}. There are more general matching models with weights attached to the ``boy'' vertices or ``girl'' vertices arriving from a known distribution or a random order \cite{mehta2013}. It would be interesting to see if our mechanisms remain most competitive in such models.

\bibliographystyle{splncs}
\bibliography{competitive}

\end{document}